\documentclass[final,3p,times]{elsarticle}
\usepackage{amsmath,amssymb,amsfonts,scalerel}
\usepackage{graphicx}
\usepackage{subcaption}
\usepackage{booktabs}
\usepackage{array}
\usepackage{float}
\usepackage{xcolor}
\usepackage[colorlinks=true,citecolor=blue,linkcolor=blue,urlcolor=blue]{hyperref}
\usepackage{setspace}
\usepackage{multirow}
\usepackage{hhline}
\usepackage{multicol}
\usepackage{adjustbox}
\usepackage{url}
\usepackage{rotating}
\usepackage{geometry}
\biboptions{sort,compress}

\begin{document}
\begin{frontmatter}

\title{Layering effect on exciton and thermoelectric characteristics}

\author[1]{Eyasu~Tadese.~M}
\ead{joshtadese@gmail.com}

\author[1]{T.E.~Ada\corref{cor1}}
\cortext[cor1]{Corresponding author:\ tewodros.eyob@aau.edu.et\ 
               (T.E.~Ada)}
               
\author[3]{ K.N.~Nigussa}
\ead{kenate.nemera@aau.edu.et}

\author[4,5]{Cecil~N.M.~Ouma}
\ead{couma@aimsric.org}

\address[1]{Department of Physics, Dilla University, P.O. Box 419, 
         Dilla,\ Ethiopia}
         
\address[3]{Department of Physics, Addis Ababa University, P.O. Box 1176, 
         Addis Ababa,\ Ethiopia} 
         
\address[4]{African Institute for Mathematical Sciences (AIMS) | Research $\&$ Innovation Centre, Kigali, Rwanda}  

\address[5]{HySA Infrastructure Center of Competence, North West University, Faculty of Engineering, P. Bag X6001, Potchefstroom 2520, South Africa}

\begin{abstract}

Ab initio calculation was performed using the $\mathrm{vdW-DF2}$ functional coupled with a modified Schr$\rm\ddot{o}$dinger equation to calculate exciton binding energy from the electron-hole wave function.  The findings indicate that hetero-layering within the same AA-stacking has a considerable effect on the band edge, altering the effective electron mass and hole distribution. The size and shape of electron and hole clouds in hetero-layered van der Waals with different kinds of neighboring layers improve intermolecular interactions, resulting in a narrower electronic gap and higher exciton binding energy.  Whenever these excitons are exposed to sunlight for an extended period of time, they begin to disintegrate, releasing hot electrons with some energy; yet, because they have more unoccupied states, they can remain at greater energy levels for longer periods. This, in turn, enhances cooling by spreading heat evenly among available states.  At higher temperatures, we noticed that some layer topologies had a greater density of state due to quasiparticles such plasmons and surface plasmon-polariton-assisted hot electron excitation. Hence, layering at a maximum plurality enhances exciton or thermoelectric capabilities. 
\end{abstract}

\end{frontmatter}

\section{Introduction\label{sec:intro}}

As the world attempts to achieve Net Zero, the transition from fossil-based to sustainable energy sources is critical. There is a rush to scale technology improvements along the clean energy value chain, and photovoltaic~(PV) systems and gadgets have gained popularity in this regard. However, the most significant impediment to these technological advancements is the efficiency of the devices~\cite{LDPRXCXNZ_2022, MFFVAL_2014}. Materials employed in the manufacturing of these devices are recognized to have a role, owing to the intrinsic properties they possess, and this has prompted research into innovative materials that will solve the shortcomings of those already in use. 

2D van der Waals heterostructures are one type of material being studied for high-efficiency solar cells. A 2D material can have exceptional properties, particularly in optics and electron transport, because to strong covalent bonding, which provides in-plane stability, van der Waals forces, which hold the stack together, and electron-hole pairing among adjacent layers~\cite{SLKWTOKT_2016}.

Interlayer electron-hole pairing occurs in a 2D van der Waals heterostructure as there is a vacuum between the bottom of the conduction band and the top of the valence band, and the intermolecular interactions between uneven electron density layers separate them. Thus, interlayer electron-hole pairing in various configurations is anticipated to have a major impact on photon absorption and optical properties in solar cells. 

At moderate temperatures, exciton-binding energy maintains electron-hole pairing until the exciton dissociates into hot electrons at high temperatures due to solar cell heating, causing hot carriers to populate the conduction band or emit light as carriers drop from conduction to valence band with the release of Helmholtz free energy, or cooling hot carriers for a longer lifetime at a higher energy level.  Hot carrier relaxation is often associated with the appearance of new unoccupied states, which distinguishes materials that increase density of state as they heat. This is often referred to as the thermoelectric effect.

Therefore, a thorough understanding of interlayer electron-hole pairing and hot electrons is critical for constructing a 2D van der Waals heterostructure in various configurations. However, such a system is computationally expensive, and density functional theory approaches based on the Beth-Salpeter Equation (BSE) are exceedingly expensive for computing band gaps and band alignment at 2D semiconductor interfaces~\cite{RFSPWKTK_2016, BKR_1981, LSOTTKS_2015}.

In this work, we used a screened hydrogen model to calculate the binding energy of excitons across various configurations. We also investigated the interlayer electron-hole pairing and hot electrons in a van der Waals heterostructure of 2H-bilayer~$\mathrm{\lbrace Mo,W\rbrace S_{2}/\lbrace Mo, W\rbrace Se_{2}}$ in various configurations of the highest symmetry point of the first Brillouin zone, which is where the lowest bound exciton is located. Thus, our research focuses on using atomistic techniques to increase solar and thermoelectric qualities as a direct strategy for reducing future $\mathrm{CO_{2}}$ emissions by replacing future fossil fuel-powered electricity sources.
 
\begin{figure*}[htbp!]
\centering
\includegraphics[scale=0.35]{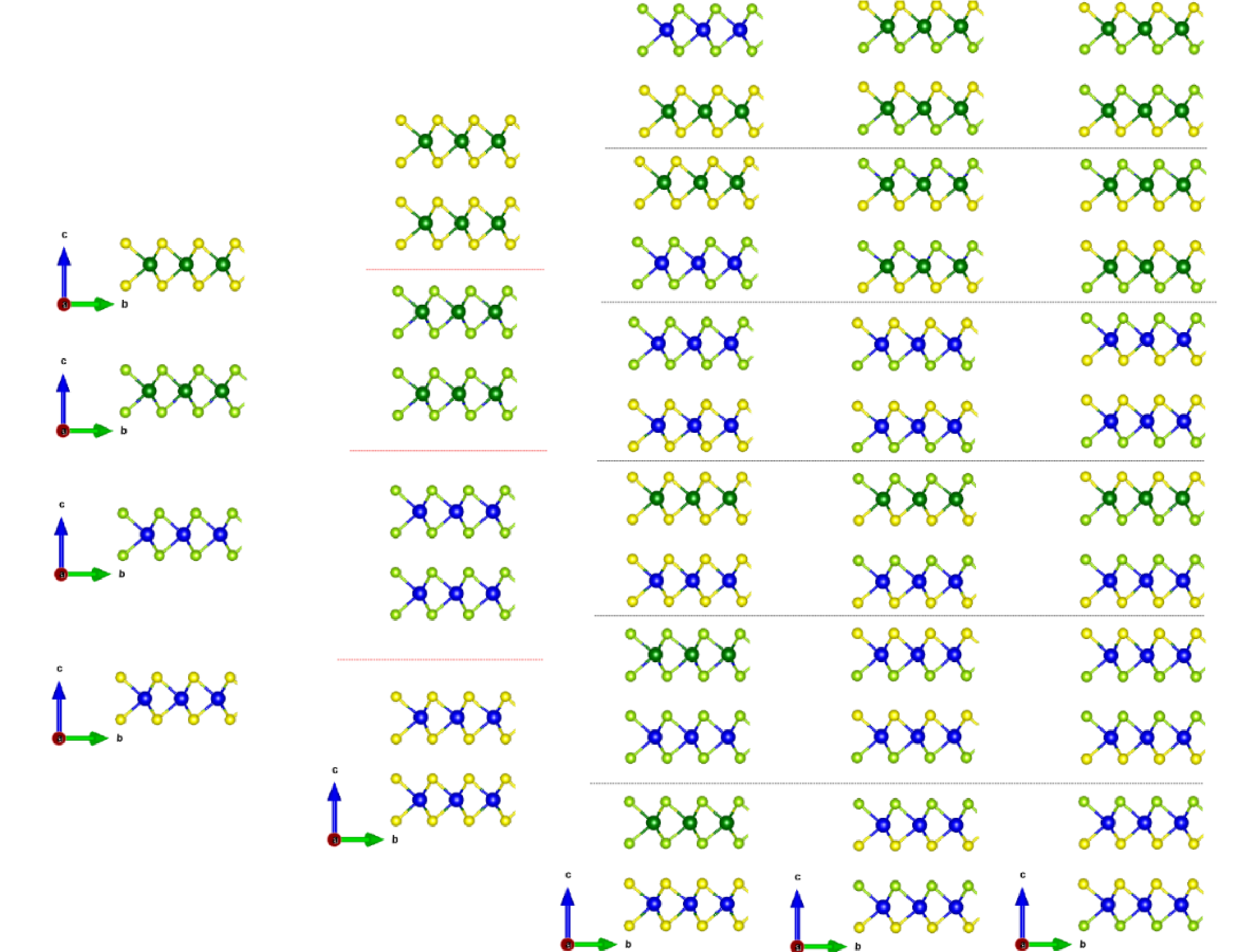}
\caption{\label{fig1} Possible combinations of layered hetrostructuring within AA-stacking consisting of $\mathrm{\{Mo, W\}S_{2}}$ and $\mathrm{\{Mo, W\}Se_{2}}$ colored by constituent elements: yellow for Sulfur and green for Selenium, dark green or blue for Molybdenum or Tungsten. Refer to the web version of this page for further information on color code interpretation.}
\end{figure*}
\section{Computational\ method\label{sec:compmeth}}
All ab-initio calculations were performed using the gpaw algorithm~\cite{JEetal2010}, and the projected augmented wave was used for Kohn-Sham self-consistent computations~\cite{MHJ2005, KS65}. The wave functions have been expanded in a plane wave cut-off energy of less than 600 eV using a double-zeta polarized atomic orbitals basis set~\cite{PB94}. The $k$-points\  within the Brillouin zone (BZ) are chosen using the Monkhorst-Pack technique~\cite{MP76}, with a 42$\times$42$\times$1 mesh size. The interactions between valence electrons, core electrons, and nuclei are approximated using projector augmented wave (paw) data sets~\cite{PB94, KJ99}. The stress tensor is used to optimize the geometry of the atomic coordinates and the degrees of freedom of unit cells~\cite{BS82, PRF39, NM85, WM91}. Force convergence conditions were set to 0.005~$\mathrm{eV/\AA}$. The generalized gradient approximation of PBE~\cite{PBE96} and a 2$\times$2$\times$1  $\mathrm{\bf{k}}$ mesh are used in geometry relaxation calculations for a 2H-heterostructure with two mono layers separated by 2.0~$\AA$ vacuum. The surface is insulated from external interaction with 10~$\AA$  vacuum along the c-axis direction.

To calculate band gaps, we used the exchange-correlation functionals $\mathrm{GLLB-SC}$ and $\mathrm{vdW-DF2}$~\cite{KOER2010, PBEsol2013}. The density of states (DOS) and group velocity, $\mathrm{\nu_{g}(\varepsilon)}$~\cite{KGNSMDRB_1990}, are calculated using a graph of $\varepsilon(k)$\ vs $k$, and DOS at a specific energy $\varepsilon$ is given as

\begin{equation}
{\label{eq1}}
\mathrm{\rho({\varepsilon})=\sum\limits_{N}w\rm_{N}{\delta}({\varepsilon}
-{\varepsilon\rm_{N}})},
\end{equation}
and,
\begin{equation}
\mathrm{\nu_{g}({\varepsilon})=\frac{k^2}{2\pi^2}\frac{1}{\rho({\varepsilon})\hbar}}
\end{equation}
where $\mathrm{N=(k,s)}$ is\ an\ occupation\ state\ corresponding\ to\ a\ $\mathrm{k}$\ point\ $\&$\ a\ spin $s$,\ $\&$\ $\mathrm{w\rm_{N}}$\ is a weight\ factor\ respectively.

To better match experimental results, we employed $\mathrm{vdW-DF2}$~\cite{HHN_1999, JKDBAM_2010, CDBKTT_2020} for layer interaction and $\mathrm{GLLB-SC}$ for London force exclusion to differentiate the van der Waals interaction force effect.

Kane's theory~\cite{NBRCAN_1975} provides the $\varepsilon(k)$\ vs $k$ relation, which analyzes the impact of band alignment at interfaces.

\begin{equation}
\mathrm{\varepsilon(k)=\frac{\hbar^{2} k^{2}}{2m^{*}_{N}}-\alpha \varepsilon(k)^{2}}
\label{eq3}
\end{equation}
Where the effective mass,  $\mathrm{m^{*}_{N}}$ and energy, $\mathrm{\varepsilon}$ are measured at the band edge, respectively and $\alpha$ is used to quantify the flattening of a parabola curve, a smaller negative value indicates that the parabolic curve is more flat. Thus, accurately estimating the effective mass from the $\mathrm{\varepsilon(k)}$ curve is crucial for correctly modeling the optical and transport properties of various structural configurations. To compute effective mass, we considered energy change, $\mathrm{\varepsilon(k)}$, along the k-point from $\mathrm{A}$ or $\mathrm{\Gamma}$ to $\mathrm{K}$ or $\mathrm{H}$, where $\mathrm{\Gamma~[0.0, 0.0, 0.0]}$,  $\mathrm{A~[0.0, 0.0, 1/2]}$, $\mathrm{K~[1/3, 1/3, 0.0]}$,  and $\mathrm{H~[1/3, 1/3, 1/2]}$ were used to calculate effective mass~($\mathrm{m_{e}~\&~ m_{h}}$) averaged over size of electrons or holes. Thus, using Eqs.~(\ref{eq3}) and (\ref{eq6}) one can calculate the average band curvature, $\mathrm{\alpha}$.

The Taylor expansion can be used to discover the $\mathrm{\varepsilon(k)}$ dispersion relation as follows: 
\begin{equation}
\mathrm{\varepsilon(k)=\varepsilon_{0}+\frac{\partial \varepsilon}{\partial k}+\frac{\partial^2 \varepsilon}{2!\partial k^2}+O(k^3)}
\label{eq4}
\end{equation}
Where $\mathrm{\varepsilon_{0}}$ is energy at $\Gamma$-point.
Then, contracting Eq.\ref{eq4} and discarding the second term, we get,
\begin{equation}
\mathrm{\varepsilon(k)=\varepsilon_{0}+\alpha |\Delta k|^2}
\label{eq5}
\end{equation}

Hence, the change in $\mathrm{\varepsilon(k)}$ curve at a symmetric path k-point divided by the change in square of the k-point path equals twice the band curvature. 
\begin{equation}
\mathrm{\frac{\hbar^2}{2m^*_{N}}=\frac{\partial^2 \varepsilon}{2!\partial k^2}=\alpha}
\label{eq6}
\end{equation}
$\mathrm{\frac{\hbar^2 k^2}{2m^*_{N}}=\alpha}$, this relation proves that materials with small masses have high mobility and long diffusion lengths~\cite{BFBKWASM_2014}.

The Beth-Salpeter Equation~(BSE) is commonly used for studying the exciton effect~\cite{BXDIJDLP_2020}, but its application to van der Waals heterostructures with a few stacked layers is computationally challenging.  The screened hydrogen model, on the other hand, reduces the exciton binding energy computation to just the reduced effective mass, $\mu$ and dielectric constant, $\epsilon_{0}$, both of which can be calculated using ab-initio methods. Thus, exciton binding energy in atomic units is stated as,

\begin{equation}
\mathrm{E_{B}^{3D}=\frac{\mu}{2\epsilon_{0}^{2}}}
\label{eq7}
\end{equation}

The screening formula for 2D dielectrics is $\epsilon(\textbf{q})=1+2\pi \beta a$, where  $\beta$ represents the material's polarizability as determined by first-principles calculations. The screening process is non local in real space, making it difficult to predict a hydrogenic model. However, one can calculate the 2D screened potential and solve the Schr$\rm\ddot{o}$dinger equation for the electron-hole wave function as,

\begin{equation}
\left[-\frac{\nabla^{2}}{2\mu}+W(\textbf{r})\right] \Psi(\textbf{r})=E_{n}\Psi(\textbf{r})
\label{eq8}
\end{equation}

$W(\bf{r})$ is a 2D convolution of the Coulomb interaction and microscopic dielectric function, $\epsilon^{-1}(\bf{r}-\bf{r'})$. This method has already been found to have good agreement with the Beth-Salpeter equation.

The effective dielectric constant is calculated by averaging $\varepsilon(\bf{q})$ over the $\bf{q}$-vectors up to $1/a$, where $a$ denotes the radius of the exciton in real space,
\begin{equation}
\epsilon_{eff}=\frac{a^{2}}{\pi}\int_{0}^{2\pi} \int_{0}^{1/a}dq \epsilon(\bf{q})
\label{eq9}
\end{equation}
Where $a$ denotes the effective Bohr radius. The Bohr radius for a 2D hydrogen atom is $a=\frac{\epsilon}{2\mu}$, and Eq.~(\ref{eq10}) requires a self-consistent solution for $\epsilon_{eff}$ given an expression for $\epsilon(\bf{q})$. In a purely 2D system, the screening is linear in q, and the Eq.~(\ref{eq9}) can be solved to obtain,

\begin{equation}
\epsilon_{eff}=\frac{1}{2}\left(1+\sqrt{1+32\pi \beta \mu/3}\right)
\label{eq10}
\end{equation}
We established that the hydrogen binding energy model in two dimensions is four times higher than in three dimensions, and it scales linearly with band gaps. 

\begin{equation}
E_{B}^{2D}=\frac{8\mu}{\left(1+\sqrt{1+32\pi \beta \mu/3}\right)^{2}}
\label{eq11}
\end{equation}

\begin{figure}[htbp!]
        \centering
        \includegraphics[scale=0.5]{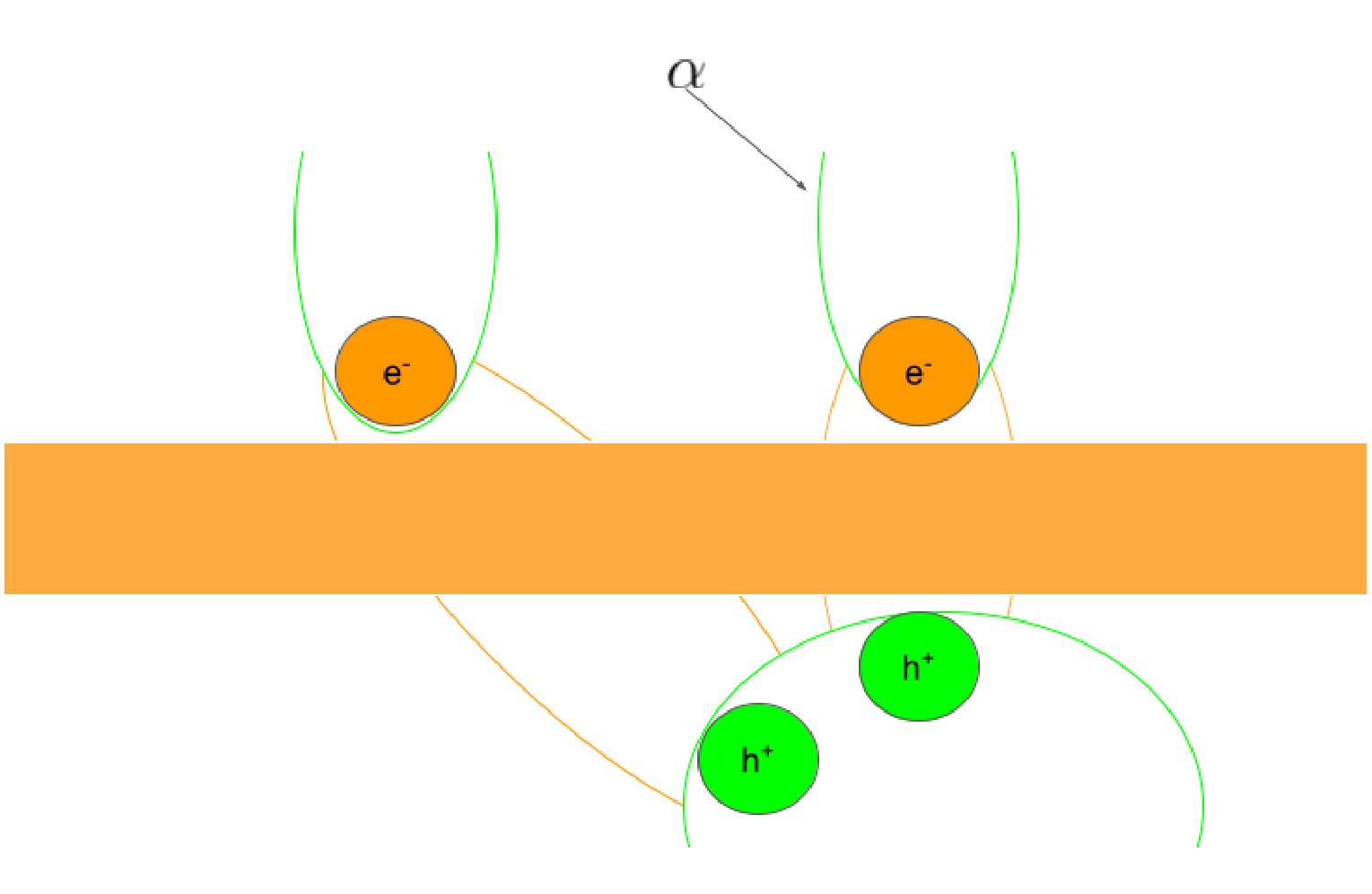}
        \caption{\label{fig2} In response to band curvature, $\mathrm{\alpha}$ the band alignment of hetro layered configuration initiates either direct or indirect electron and hole binding.}
\end{figure}

\begin{figure}[htbp!]
        \centering
        \includegraphics[scale=0.85]{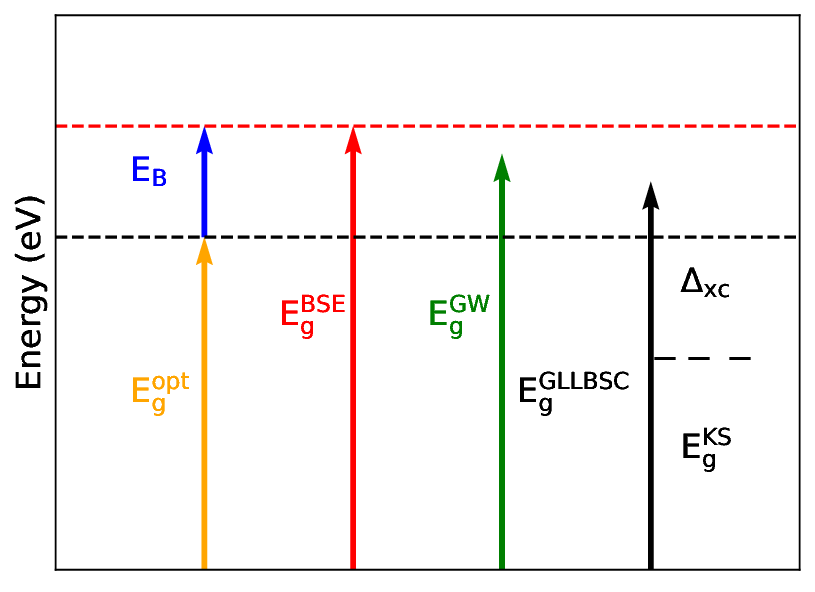}
        \caption{\label{fig3} The fundamental gap, $\mathrm{E_{g}^{BSE}}$ can be calculated as the sum of the optical gap, $\mathrm{E_{g}^{opt}}$ and exciton binding, $\mathrm{E_{B}}$, whereas the optical gap is also composed of the electronic gap and exciton binding energy; however, the electronic gap, $\mathrm{E_{g}^{KS}}$ is always smaller than the optical gap. For interpretation of color code references, refer to the web version of this article.}
\end{figure}
\section{Results and discussion\label{sec:concc}}

We used the UnitCellFilter optimizer, which is provided in the gpaw packages, to optimize the equilibrium lattice constants for each of the configurations shown in~Fig~\ref{fig1}. The estimated parameters correspond to the experimental values presented in Ref~\cite{KJTSZJLJWJ_2013, RAYMES_2006, HZ2004}. The predicted effective masses for various configurations are commensurate with those reported in the literature~\cite{CTLWL_2012, PRDTWKT_2019, LSMHFBKK_2018}. Van der Waals interactions between layers have a significant impact on a material's effective mass and band curvatures. In this work, we calculated and analyzed effective mass using the $\mathrm{GLLB-SC}$ exchange correlation. We found that the value is underestimated since the correlation ignores the London force. As a result, using a more improved van der Waals interaction exchange correlation functional, such as $\mathrm{vdW-DF2}$, yields the best results. 

\subsection{Excitonic Property}
Electrons and holes interact more strongly in two-dimensional electrical systems. These two quasiparticels come together with excitonic binding energy and can be calculated using band structure, which usually involves using effective mass and a dielectric screen. The polarizability of a material controls its dielectric screening, or the rate at which an electron cloud surrounding a molecule is deformed by an external electric field, resulting in a temporary dipole. It is crucial to quantify the strength of intermolecular interactions.The strength of the interaction determines exciton binding energy. Fig.~\ref{fig2} depicts how excitonic effects bind at either different or identical k-points, resulting in variations in the exciton energy. Electrons can excite either directly or indirectly, depending on their band orientation.

The Bethe-Salpeter equation~(BSE) has been demonstrated to be reliable in determining the quasi-particle band gap. The Coulomb term defines the screened electron-hole interaction that produces excitons and is used to properly compute the optical gap. Fig.~\ref{fig3} shows that a quasi-particle band gap calculated using the Bethe Salpeter Equation equals the optical gap plus the exciton binding energy. Our work used the hydrogen model to calculate the exciton binding energy of bilayer $\mathrm{2H-MoS_{2}}$, which is in agreement with the BSE value reported in Ref~\cite{OTLSRFTKS_2016}. One can note that $\mathrm{E_{B}+E_{g}^{opt}=E_{g}^{BSE}}$, exciton binding energy and optical gap are often difficult to distinguish due to a lack of a distinct line between photoabsorption and resonant excitation energy. Thus, the estimated sum values of exciton binding energy and electronic gap are  similar to the experimental values as published in Refs~\cite{HFOTTK_2013, JVJDDTGS_2020, Bhatnagar2022, Philipp_2013, Liu_2015}. 

\begin{figure}[htbp!]
        \centering
        \includegraphics[scale=0.5]{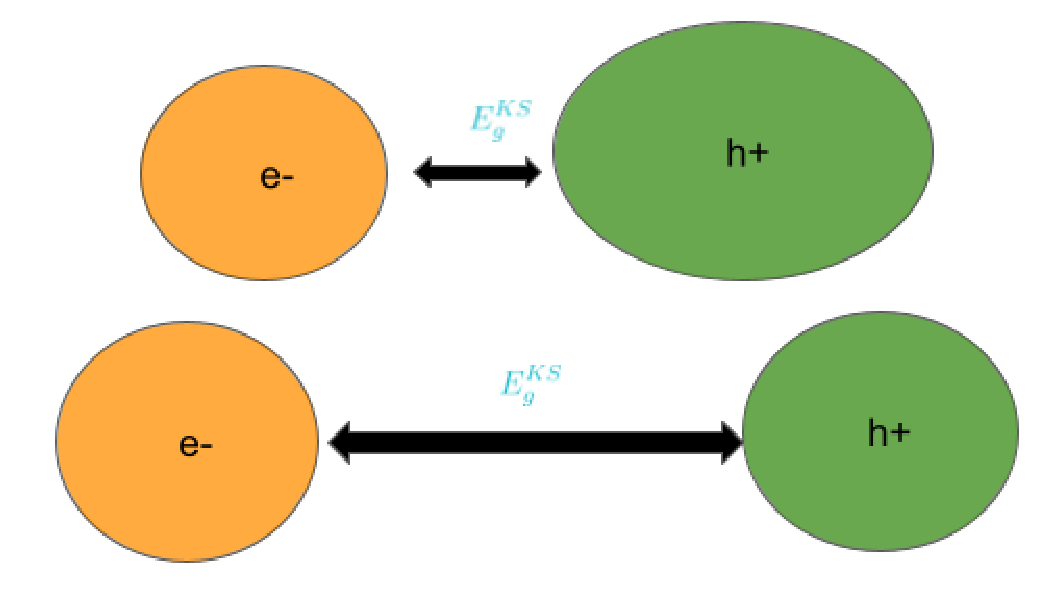}
        \caption{\label{fig8} The deformed electron cloud is primarily influenced by unequal electron and hole sizes, which in turn alter the interaction range in between.}
\end{figure}

Table~\ref{tabMono} demonstrates that the four monolayer dichalcogenides have polarizability values similar to those reported in Ref~\cite{KIKH_2015, BTHMRD_2013}, but the exciton binding energy varies due to the band alignment effect, Fig~.\ref{fig8} shows how the size and amount of effective electron masses and holes affect the dispersion force, exciton binding energy, the electronic band gap, and thermoelectric property.  It is worth noting that $\mathrm{WSe_{2}}$ has a significantly higher density of electrons and holes, resulting in higher exciton binding energy due to stronger interaction, which is greatly aided by hybridization of valence electrons of constituent atoms. However, bilayer $\mathrm{WSe_{2}}$ possesses weaker intermolecular interactions due to the extremely strong confined monolayer interaatomic interaction beneath it. As a result, layer interaction is lowered by approximately 25$\%$ and could have been strengthened by providing more accessible states for bonding due to hybridization.

\begin{figure*}[htbp!]
\begin{center}
\begin{subfigure}[t]{0.5\textwidth}
\includegraphics[width=\textwidth]{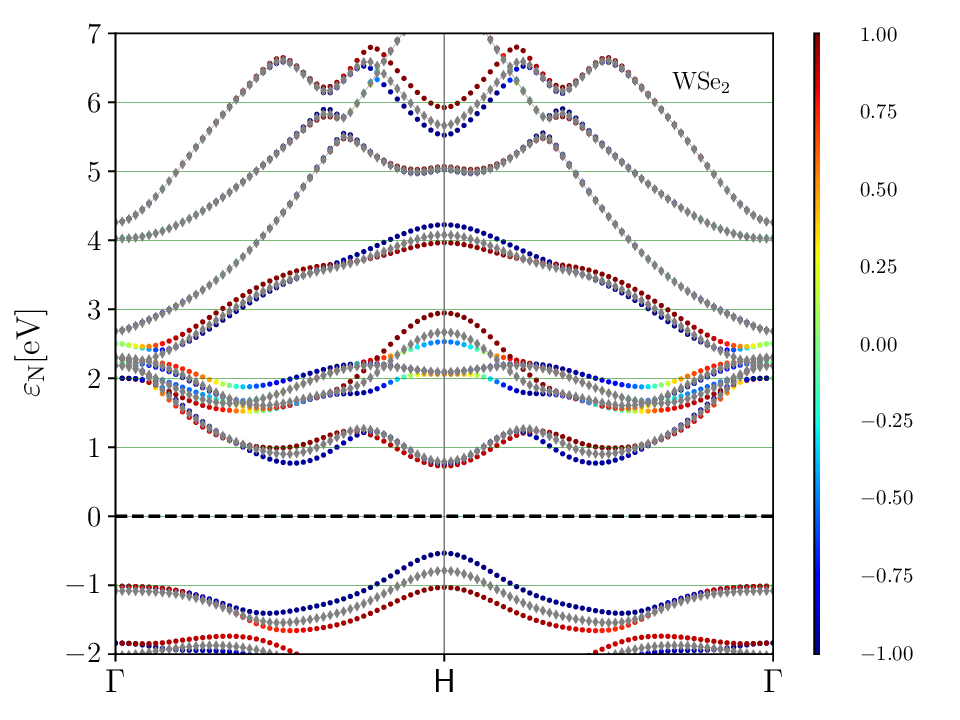}
\end{subfigure}%
\begin{subfigure}[t]{0.5\textwidth}
\centering
\includegraphics[width=\textwidth]{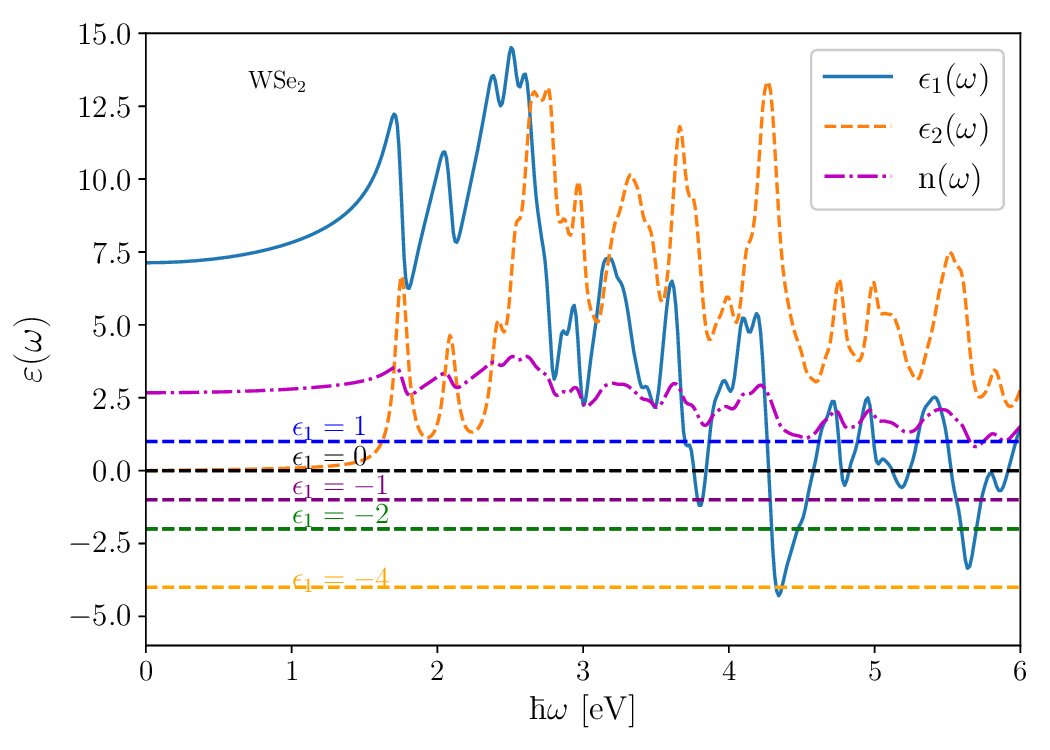}
\end{subfigure}
\begin{subfigure}[t]{0.5\textwidth}
\includegraphics[width=\textwidth]{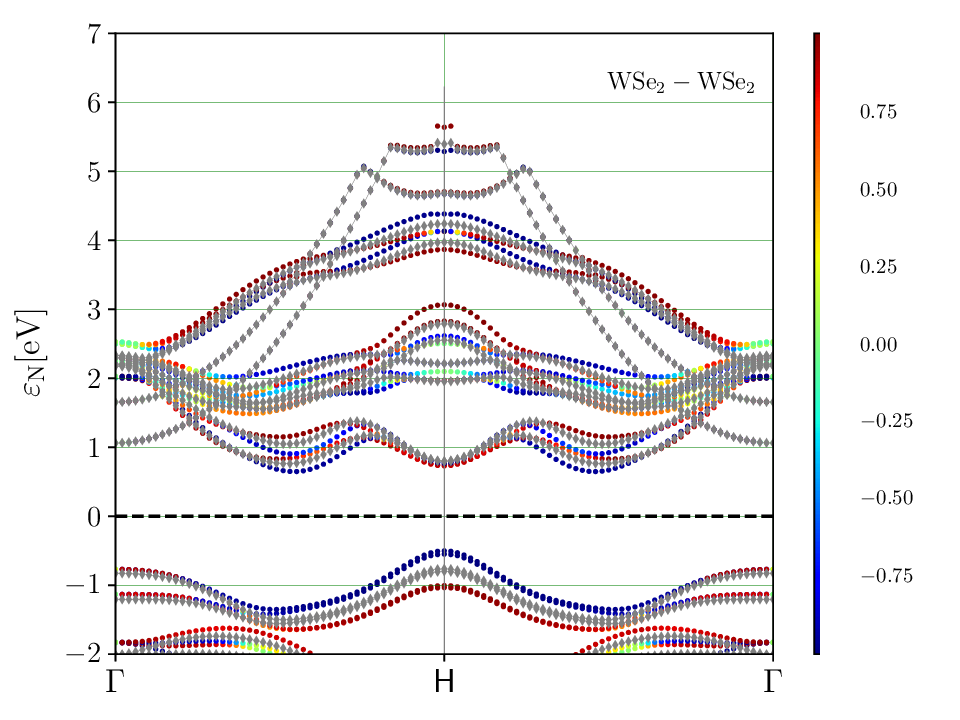}
\end{subfigure}%
\begin{subfigure}[t]{0.5\textwidth}
\centering
\includegraphics[width=\textwidth]{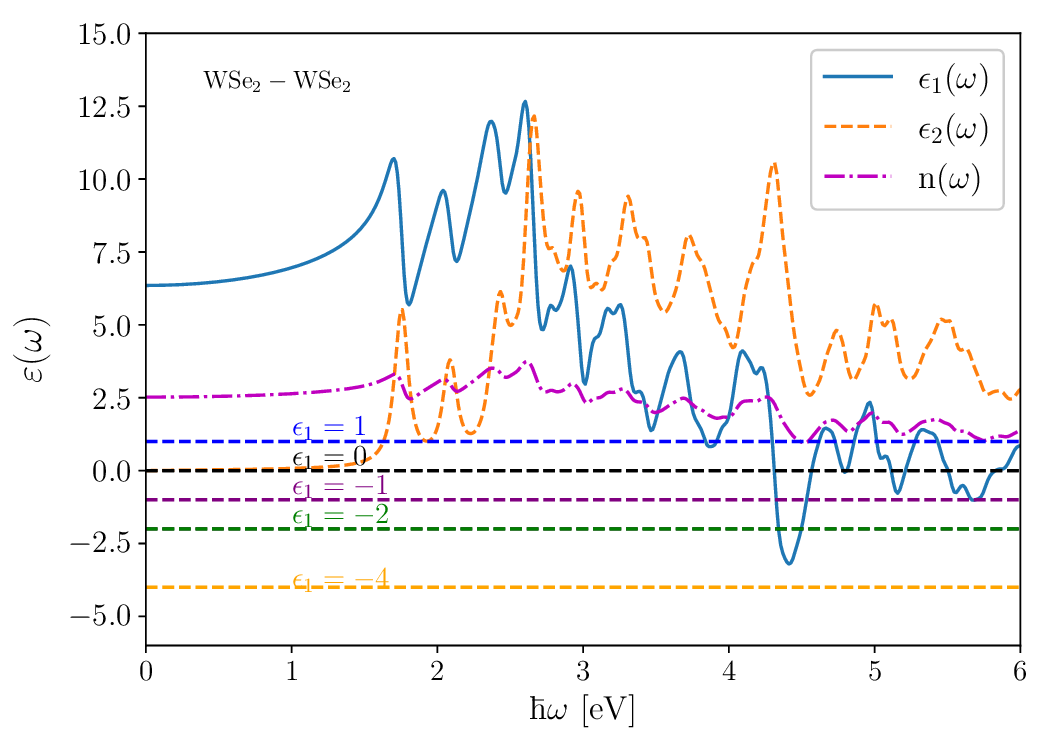}
\end{subfigure}

\caption{\label{figg23at1} Calculation of monolayer the $\varepsilon(k)$ relation to hybridization intensity in comparison to bilayer is shown on the band structures, and gray colored marked as non-hybridized band, $\varepsilon_{N}$ on the left and the dielectric function, $\mathrm{\epsilon(\omega)}$ $\&$ index of refraction, $\mathrm{n(\omega)}$ on the right.}
\end{center}
\end{figure*}

Table~\ref{tab1a} shows that layering of monolayers does not always improve layer interaction; consequently, there must be a sufficient number of electrons or holes, or hybridization should offer more accessible states. In return, band alignment significantly reduces effective mass, holes, exciton energy, and polarizability, and is often proportionate to the amount of electrons in a cloud.
Monolayers have fewer states than bilayers, however hybridization-induced carrier state evolution is possible, as evidenced by scaled energy levels on the bandstructure and surface plasmon polariton oscillation with $\mathrm{\frac{\omega_{p}^{2}}{\sqrt{5}}}$ on Fig~\ref{figg23at1}.

Table~\ref{tab2a} shows that putting a monolayer of $\mathrm{WSe_{2}}$ in a heterolayer made up of mono layers of the same type affects both excitonic and thermoelectric properties by limiting the number of accessible carriers. In contrast, stacking $\mathrm{MoS_{2}}$  with other types has a positive effect.

Table~\ref{tab3a} and \ref{tab4a} show that hetro layered van der Waals configurations using different surrounding layers boost intermolecular forces, resulting in a narrower electronic gap and higher exciton binding energy. In contrast, the same kind of layering receives less interaction.

\begin{table*}[htbp!]
\setlength{\tabcolsep}{6.0mm}
\renewcommand{\arraystretch}{1.4}
\centering
\small
\caption{ The ab-initio band structure computation yielded the reduced effective mass, $\mu$, derived from effective electron and hole masses. The exciton binding energy, $\mathrm{E_{B}}$, is calculated using the screened hydrogen model equations Eqs. ~(\ref{eq7}), (\ref{eq8}) and (\ref{eq11}), taking into account the material's polarizability, $\beta$ and dielectric constant, $\mathrm{\epsilon_{0}}$. The electronic band gap has also been computed using $\mathrm{vdW-DF2}$ and $\mathrm{GLLB-SC}$ hybrid functionals along the electron transition routes for mono layer dichalcogenides.}
\label{tabMono}
\centering
\resizebox{\textwidth}{!}{%
\begin{tabular}{c c c c c c c c c c c} 
\hline\hline
System&$\mathrm{~m^{*}}$&$\mu$&$\mathrm{E_{B}}$&$\mathrm{\beta~[\AA]}$&$\mathrm{\epsilon(\omega=0)}$&$\mathrm{vdW-DF2\left(E^{KS}_{g}\right)}$&$\mathrm{GLLB-SC\left(E^{KS}_{g}\right)}$ &$\mathrm{Transition,~V~\rightarrow~C}$&\\  \hline
&$\mathrm{m_{h}=-0.34}$&&&&&&&&\\
~$\mathrm{MoS_{2}}$&&0.13&0.25&5.76&9.8&1.58&1.65&{$\mathrm{A~\rightarrow~K}$}&\\ 
&$\mathrm{m_{e}=0.20}$&&&&&&&&\\ 
\hline
&$\mathrm{m_{h}=-0.27}$&&&&&&&\\
~$\mathrm{MoSe_{2}}$&&0.11&0.21&6.61&8.0&1.45&1.45&{$\mathrm{H~\rightarrow~K}$}&\\ 
&$\mathrm{m_{e}=0.19}$&&&&&&&&\\ 
\hline
&$\mathrm{m_{h}=-0.54}$&&&&&&&&\\
~$\mathrm{WSe_{2}}$&&0.22&0.42&5.55&7.5&1.58&1.54&{$\mathrm{H~\rightarrow~H}$}&\\ 
&$\mathrm{m_{e}=0.38}$&&&&&&&&\\
\hline
&$\mathrm{m_{h}=-0.27}$&&&&&&&\\
~$\mathrm{WS_{2}}$&&0.12&0.24&4.74&6.7&1.86&1.81&{$\mathrm{H~\rightarrow~K}$}&\\ 
&$\mathrm{m_{e}=0.23}$&&&&&&&\\ 
\hline\hline
\end{tabular}}
\end{table*}

\begin{table*}[htbp!]
\setlength{\tabcolsep}{6.0mm}
\renewcommand{\arraystretch}{1.4}
\centering
\small
\caption{The ab-initio band structure computation yielded the reduced effective mass, $\mu$, derived from effective electron and hole masses. The exciton binding energy, $\mathrm{E_{B}}$, is calculated using the screened hydrogen model equations Eqs. ~(\ref{eq7}), (\ref{eq8}) and (\ref{eq11}), taking into account the material's polarizability, $\beta$ and dielectric constant, $\mathrm{\epsilon_{0}}$. The electronic band gap has been also computed using $\mathrm{vdW-DF2}$ and $\mathrm{GLLB-SC}$ hybrid functionals along the electron transition routes for homo layered van der Waals structures.}
\label{tab1a}
\centering
\resizebox{\textwidth}{!}{%
\begin{tabular}{c c c c c c c c c c c} 
\hline\hline
System&$\mathrm{~m^{*}}$&$\mu$&$\mathrm{E_{B}}$&$\mathrm{\beta~[\AA]}$&$\mathrm{\epsilon(\omega=0)}$&$\mathrm{vdW-DF2\left(E^{KS}_{g}\right)}$&$\mathrm{GLLB-SC\left(E^{KS}_{g}\right)}$ &$\mathrm{Transition,~V~\rightarrow~C}$&\\  \hline
&$\mathrm{m_{h}=-0.50}$&&&&&&&&\\
~$\mathrm{MoS_{2}-MoS_{2}}$&&0.20&0.37&11.36&6.3&1.46&1.54&{$\mathrm{A~\rightarrow~K}$}&\\ 
&$\mathrm{m_{e}=0.34}$&&&&&&&&\\ 
\hline
&$\mathrm{m_{h}=-0.95}$&&&&&&&\\
~$\mathrm{MoSe_{2}-MoSe_{2}}$&&0.41&0.69&13.11&7.1&1.43&1.44&{$\mathrm{H~\rightarrow~H}$}&\\ 
&$\mathrm{m_{e}=0.71}$&&&&&&&&\\ 
\hline
&$\mathrm{m_{h}=-0.21}$&&&&&&&&\\
~$\mathrm{WSe_{2}-WSe_{2}}$&&0.084&0.16&12.37&6.6&1.52&1.52&{$\mathrm{H~\rightarrow~\frac{1}{2}H}$}&\\ 
&$\mathrm{m_{e}=0.14}$&&&&&&&&\\
\hline
&$\mathrm{m_{h}=-0.76}$&&&&&&&\\
~$\mathrm{WS_{2}-WS_{2}}$&&0.29&0.52&10.67&5.9&1.65&1.70&{$\mathrm{\frac{2}{3}A~\rightarrow~H}$}&\\ 
&$\mathrm{m_{e}=0.46}$&&&&&&&\\ 
\hline\hline
\end{tabular}}
\end{table*}

\begin{table*}[htbp!]
\setlength{\tabcolsep}{6.0mm}
\renewcommand{\arraystretch}{1.4}
\centering
\small
\caption{The ab initio band structure computation yielded the reduced effective mass, $\mu$, derived from effective electron and hole masses. The exciton binding energy, $\mathrm{E_{B}}$, is calculated using the screened hydrogen model equations Eqs. ~(\ref{eq7}), (\ref{eq8}) and (\ref{eq11}), taking into account the material's polarizability, $\beta$ and dielectric constant, $\mathrm{\epsilon_{0}}$. The electronic band gap has been also computed using $\mathrm{vdW-DF2}$ and $\mathrm{GLLB-SC}$ hybrid functionals along the electron transition routes for hetro layered van der Waals structures.}
\label{tab2a}
\centering
\resizebox{\textwidth}{!}{%
\begin{tabular}{c c c c c c c c c c c} 
\hline\hline
System&$\mathrm{~m^{*}}$&$\mu$&$\mathrm{E_{B}}$&$\mathrm{\beta~[\AA]}$&$\mathrm{\epsilon(\omega=0)}$&$\mathrm{vdW-DF2\left(E^{KS}_{g}\right)}$&$\mathrm{GLLB-SC\left(E^{KS}_{g}\right)}$ &$\mathrm{Transition,~V~\rightarrow~C}$&\\  \hline
&$\mathrm{m_{h}=-1.00}$&&&&&&&&\\
~$\mathrm{MoSe_{2}-MoS_{2}}$&&0.35&0.61&12.26&6.8&0.64&0.71&{$\mathrm{H~\rightarrow~H}$}&\\ 
&$\mathrm{m_{e}=0.53}$&&&&&&&\\ 
\hline
&$\mathrm{m_{h}=-0.74}$&&&&&&&\\
~$\mathrm{MoS_{2}-WSe_{2}}$&&0.28&0.50&11.33&6.7&0.42&0.47&{$\mathrm{K~\rightarrow~K+\frac{2}{3}A}$}&\\ 
&$\mathrm{m_{e}=0.44}$&&&&&&&\\ \hline
&$\mathrm{m_{h}=-0.81}$&&&&&&&\\
~$\mathrm{WSe_{2}-WS_{2}}$&&0.31&0.55&10.27&6.3&0.79&0.83&{$\mathrm{K~\rightarrow~H}$}&\\ 
&$\mathrm{m_{e}=0.51}$&&&&&&&&\\
\hline
&$\mathrm{m_{h}=-0.84}$&&&&&&&&\\
~$\mathrm{MoS_{2}-WS_{2}}$&&0.33&0.58&11.00&5.9&1.34&1.41&{$\mathrm{\frac{2}{3}A~\rightarrow~H}$}&\\ 
&$\mathrm{m_{e}=0.56}$&&&&&&&&\\ 
\hline
&$\mathrm{m_{h}=-0.37}$&&&&&&&&\\
~$\mathrm{MoSe_{2}-WSe_{2}}$&&0.16&0.30&12.10&7.0&1.23&1.22&{$\mathrm{K~\rightarrow~K}$}&\\ 
&$\mathrm{m_{e}=0.27}$&&&&&&&&\\ 

\hline
&$\mathrm{m_{h}=-1.03}$&&&&&&&\\
~$\mathrm{MoSe_{2}-WS_{2}}$&&0.29&0.52&11.81&6.7&1.02&1.09&{$\mathrm{K+\frac{1}{3}A~\rightarrow~K+\frac{2}{3}A}$}&\\ 
&$\mathrm{m_{e}=0.41}$&&&&&&&&\\ 
\hline\hline
\end{tabular}}
\end{table*}
\begin{table*}[htbp!]
\setlength{\tabcolsep}{6.0mm}
\renewcommand{\arraystretch}{1.4}
\centering
\small
\caption{The ab initio band structure computation yielded the reduced effective mass, $\mu$, derived from effective electron and hole masses. The exciton binding energy, $\mathrm{E_{B}}$, is calculated using the screened hydrogen model equations Eqs. ~(\ref{eq7}), (\ref{eq8}) and (\ref{eq11}), taking into account the material's polarizability, $\beta$ and dielectric constant, $\mathrm{\epsilon_{0}}$. The electronic band gap has been also computed using $\mathrm{vdW-DF2}$ and $\mathrm{GLLB-SC}$ hybrid functionals along the electron transition routes for hetro layered van der Waals structures with different kinds of surrounding layer.}
\label{tab3a}
\centering
\resizebox{\textwidth}{!}{%
\begin{tabular}{c c c c c c c c c c c} 
\hline\hline
System&$\mathrm{~m^{*}}$&$\mu$&$\mathrm{E_{B}}$&$\mathrm{\beta~[\AA]}$&$\mathrm{\epsilon(\omega=0)}$&$\mathrm{vdW-DF2\left(E^{KS}_{g}\right)}$&$\mathrm{GLLB-SC\left(E^{KS}_{g}\right)}$ &$\mathrm{Transition,~V~\rightarrow~C}$&\\  \hline

&$\mathrm{m_{h}=-0.50}$&&&&&&&&\\
~$\mathrm{MoSSe-MoSSe}$&&0.19&0.35&12.23&6.8&0.87&1.00&{$\mathrm{\Gamma~\rightarrow~K}$}&\\ 
&$\mathrm{m_{e}=0.31}$&&&&&&&&\\ 
\hline
&$\mathrm{m_{h}=-0.50}$&&&&&&&&\\
~$\mathrm{MoSeS-MoSeS}$&&0.19&0.35&12.23&6.8&0.87&1.00&{$\mathrm{\Gamma~\rightarrow~K}$}&\\ 
&$\mathrm{m_{e}=0.31}$&&&&&&&&\\ 
\hline
&$\mathrm{m_{h}=-0.96}$&&&&&&&&\\
~$\mathrm{MoSSe-WSSe}$&&0.21&0.39&11.86&6.6&1.24&1.36&{$\mathrm{\Gamma~\rightarrow~K}$}&\\ 
&$\mathrm{m_{e}=0.27}$&&&&&&&&\\
\hline
&$\mathrm{m_{h}=-1.02}$&&&&&&&\\
~$\mathrm{MoSeS-WSeS}$&&0.36&0.63&11.86&6.7&0.71&0.79&{$\mathrm{\Gamma~\rightarrow~H}$}&\\ 
&$\mathrm{m_{e}=0.55}$&&&&&&&\\ 
\hline
&$\mathrm{m_{h}=-0.60}$&&&&&&&&\\
~$\mathrm{WSSe-WSSe}$&&0.24&0.44&11.48&6.3&1.09&1.17&{$\mathrm{\Gamma~\rightarrow~H}$}&\\ 
&$\mathrm{m_{e}=0.40}$&&&&&&&&\\ 
\hline
&$\mathrm{m_{h}=-1.01}$&&&&&&&&\\
~$\mathrm{WSeS-WSeS}$&&0.36&0.63&11.48&6.6&1.09&1.17&{$\mathrm{\Gamma~\rightarrow~H}$}&\\ 
&$\mathrm{m_{e}=0.55}$&&&&&&&&\\ 
\hline\hline
\end{tabular}}
\end{table*}
\begin{table*}[htbp!]
\setlength{\tabcolsep}{6.0mm}
\renewcommand{\arraystretch}{1.4}
\centering
\small
\caption{The ab initio band structure computation yielded the reduced effective mass, $\mu$, derived from effective electron and hole masses. The exciton binding energy, $\mathrm{E_{B}}$, is calculated using the screened hydrogen model equations Eqs. ~(\ref{eq7}), (\ref{eq8}) and (\ref{eq11}), taking into account the material's polarizability, $\beta$ and dielectric constant, $\mathrm{\epsilon_{0}}$. The electronic band gap has been also computed using $\mathrm{vdW-DF2}$ and $\mathrm{GLLB-SC}$ hybrid functionals along the electron transition routes for van der Waals structures with same kinds of surrounding layer.}
\label{tab4a}
\centering
\resizebox{\textwidth}{!}{%
\begin{tabular}{c c c c c c c c c c c} 
\hline\hline
System&$\mathrm{~m^{*}}$&$\mu$&$\mathrm{E_{B}}$&$\mathrm{\beta~[\AA]}$&$\mathrm{\epsilon(\omega=0)}$&$\mathrm{vdW-DF2\left(E^{KS}_{g}\right)}$&$\mathrm{GLLB-SC\left(E^{KS}_{g}\right)}$ &$\mathrm{Transition,~V~\rightarrow~C}$&\\  \hline

&$\mathrm{m_{h}=-0.20}$&&&&&&&&\\
~$\mathrm{MoSSe-MoSeS}$&&0.13&0.25&12.23&6.8&1.39&1.47&{$\mathrm{\Gamma~\rightarrow~\frac{1}{2}H}$}&\\ 
&$\mathrm{m_{e}=0.39}$&&&&&&&&\\ 
\hline

&$\mathrm{m_{h}=-0.71}$&&&&&&&&\\
~$\mathrm{MoSeS-MoSSe}$&&0.23&0.42&12.23&7.0&1.25&1.35&{$\mathrm{\Gamma~\rightarrow~K}$}&\\ 
&$\mathrm{m_{e}=0.33}$&&&&&&&&\\ 

\hline
&$\mathrm{m_{h}=-1.01}$&&&&&&&&\\
~$\mathrm{MoSeS-WSSe}$&&0.36&0.63&11.86&6.8&1.10&1.22&{$\mathrm{\Gamma~\rightarrow~H}$}&\\ 
&$\mathrm{m_{e}=0.57}$&&&&&&&&\\
\hline
&$\mathrm{m_{h}=-1.12}$&&&&&&&&\\
~$\mathrm{MoSSe-WSeS}$&&0.38&0.66&11.86&6.4&1.28&1.34&{$\mathrm{\Gamma~\rightarrow~H}$}&\\ 
&$\mathrm{m_{e}=0.57}$&&&&&&&&\\ 
\hline
&$\mathrm{m_{h}=-0.25}$&&&&&&&&\\
~$\mathrm{WSSe-WSeS}$&&0.09&0.17&11.48&6.8&1.53&1.63&{$\mathrm{\Gamma~\rightarrow~\frac{1}{2}K+\frac{1}{2}A}$}&\\ 
&$\mathrm{m_{e}=0.13}$&&&&&&&&\\ 
\hline
&$\mathrm{m_{h}=-0.44}$&&&&&&&&\\
~$\mathrm{WSeS-WSSe}$&&0.15&0.28&11.48&6.0&1.47&1.53&{$\mathrm{\Gamma~\rightarrow~K}$}&\\ 
&$\mathrm{m_{e}=0.23}$&&&&&&&&\\ 
\hline\hline
\end{tabular}}
\end{table*}

\subsection{Thermoelectric Properties}
In thermoelectric and solar materials, the dielectric constant determines the rate of recombination losses in the absorbing layer, with a greater dielectric constant resulting in a slower recombination rate and thus higher efficiency. Material dielectric properties are influenced by several factors, especially frequency, temperature, and chemical composition. A material's dielectric response can be expressed as,

\begin{equation}
\epsilon(\omega)=\epsilon_{1}(\omega)+i\epsilon_{2}(\omega)
\end{equation}
where, $\epsilon_{1}(\omega)$ and $\epsilon_{2}(\omega)$ are the real and imaginary part of the dielectric function respectively. 

The real part of the dielectric constant, $\epsilon_{1}(\omega)$, retains energy through polarization, while the imaginary part, $\epsilon_{2}(\omega)$, dissipates energy.~Organic semiconductors with low dielectric constants,~$\mathrm{\epsilon_{0}\approx 3}$ have higher charge carrier interactions than inorganic semiconductors with high dielectric constants~$\mathrm{\epsilon_{0}>10}$~\cite{RSMAMP_2018}.

\begin{figure*}[htp!]
\begin{subfigure}[t]{0.5\textwidth}
\centering
\includegraphics[width=\textwidth]{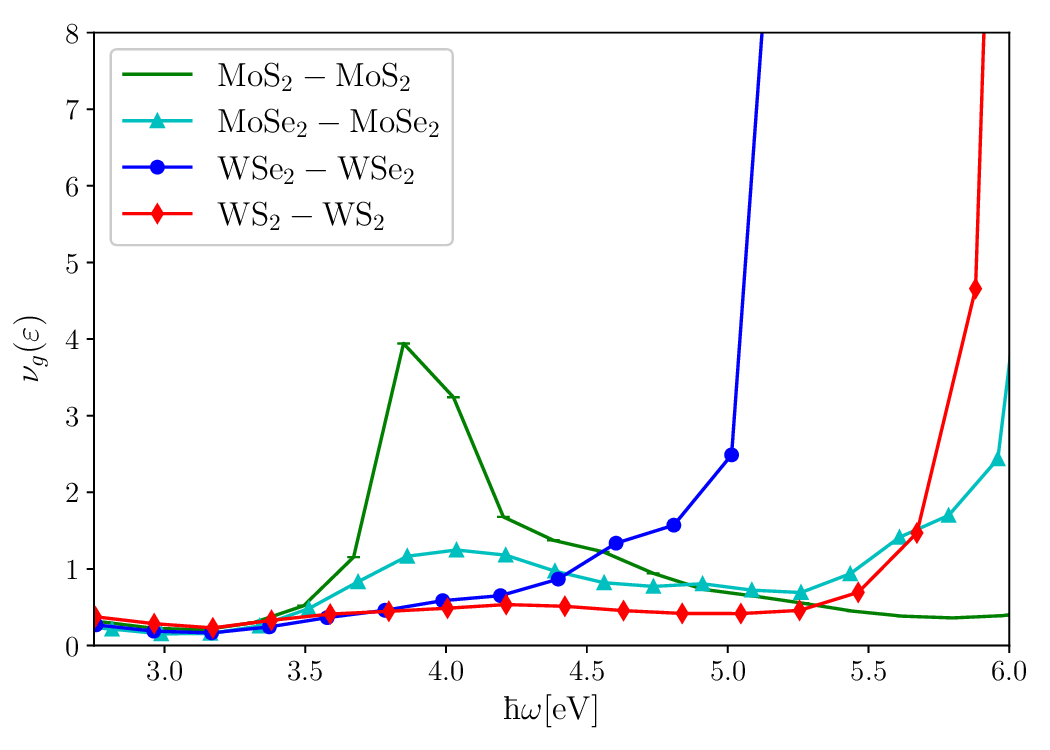}
\end{subfigure}%
\hspace{0.01\textwidth}
\begin{subfigure}[t]{0.5\textwidth}
\centering
\includegraphics[width=\textwidth]{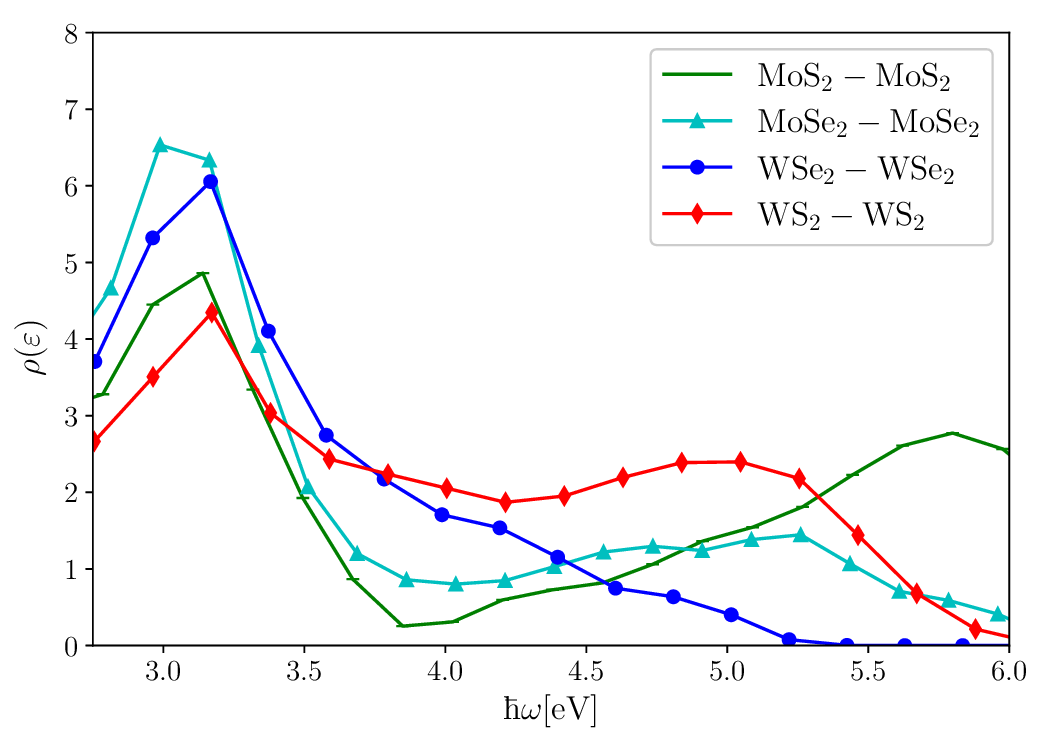}
\end{subfigure}
\caption{\label{fig1at1} Hot electron velocity, $\mathrm{\nu_{g}(\varepsilon)}$ increases in response to energy changes in their respective home structured bilayer configurations, with an evident relationship to hot electron density, $\rho(\varepsilon)$ on the right.} 
\begin{subfigure}[t]{0.5\textwidth}
\centering
\includegraphics[width=\textwidth]{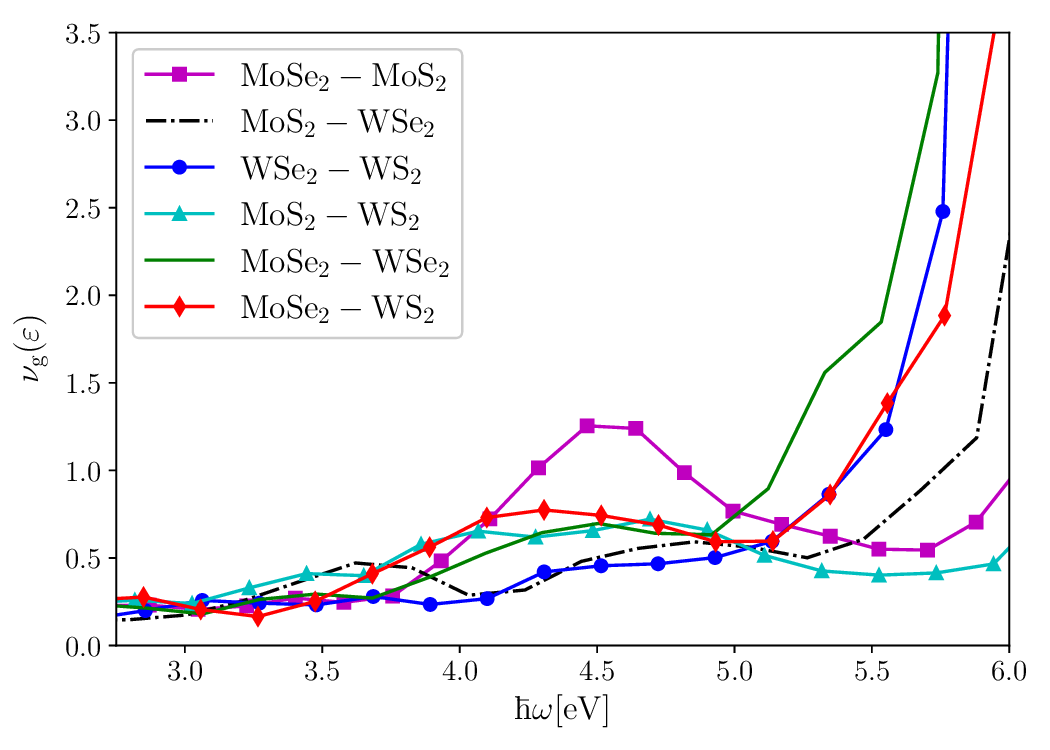}
\end{subfigure}%
\hspace{0.01\textwidth}
\begin{subfigure}[t]{0.5\textwidth}
\centering
\includegraphics[width=\textwidth]{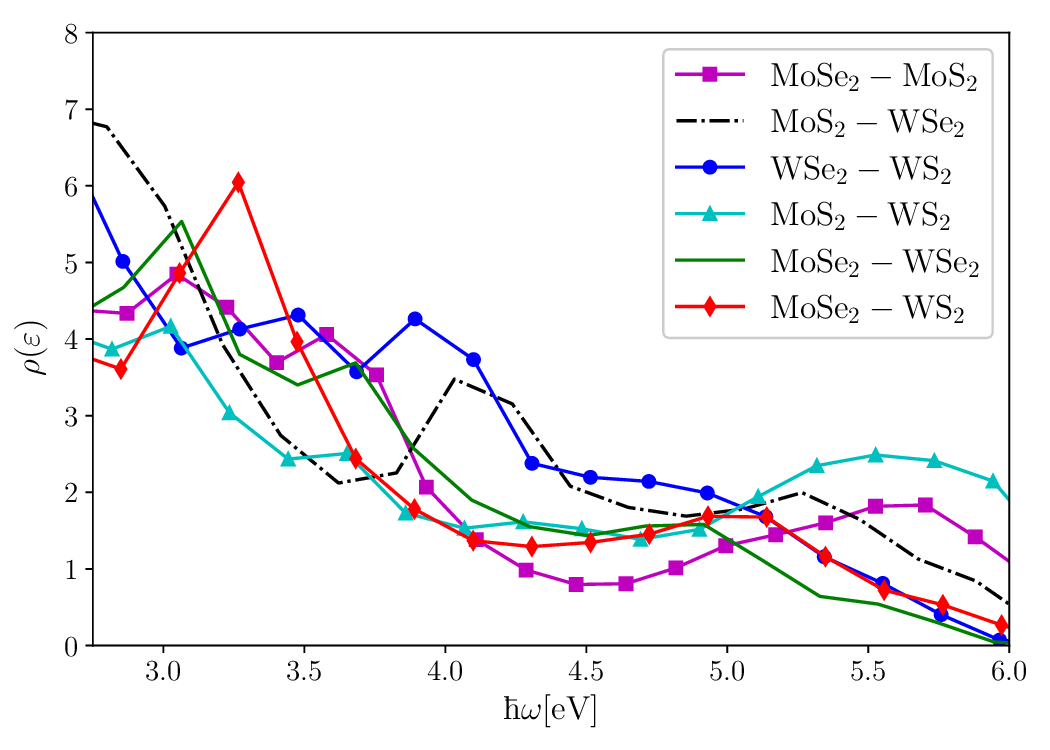}
\end{subfigure}
\caption{\label{fig1t2} Hot electron velocity, $\mathrm{\nu_{g}(\varepsilon)}$ increases in response to energy changes in their respective hetro structured bilayered configurations, with an evident relationship to hot electron density, $\rho(\varepsilon)$ on the right.} 
\end{figure*}

The density of states in a thermoelectric material increases with temperature, allowing for more unoccupied states at maximum temperatures. Thus, low thermal conductivity, achieved through density of state design or nanostructure modification, is required for the building of effective thermoelectric materials~\cite{Jiang2021, PYLAWHSJ_2012, FDK_1976, RF_2007}. Figs.~\ref{fig1at1} and \ref{fig1t2} show that in order to lower the temperature of a thermoelectric material, more hot electrons must be spread across unoccupied states. Thus, hetrostructured layer, $\mathrm{MoS_{2}-WSe_{2}}$, $\mathrm{MoSe_{2}-MoS_{2}}$, and $\mathrm{MoS_{2}-WS_{2}}$ designs perform significantly better in heat transfer than home bilayered $\mathrm{WS_{2}}$ $\&$ $\mathrm{MoS_{2}}$ arrangements due to the availability of more vacant states in respective order.

Plasmonic materials can induce considerable DOS at certain frequencies due to the presence of high-momentum states that are absent in regular materials. The Drude model defines the dielectric permittivities of plasmonic materials.

\begin{equation}
\epsilon_{m}(\omega)=\epsilon_{\infty}-\frac{\omega^{2}_{p}}{\omega^{2}+i\gamma \omega}
\end{equation}
where $\epsilon_{\infty}$ is the high-frequency permittivity limit, $\omega^{2}_{p}$
is the plasma frequency and $\gamma$ is the electron collision frequency~\cite{SvBHGG_2013}.

\begin{figure*}[htbp!]
\begin{subfigure}[t]{0.5\textwidth}
\centering
\includegraphics[width=\textwidth]{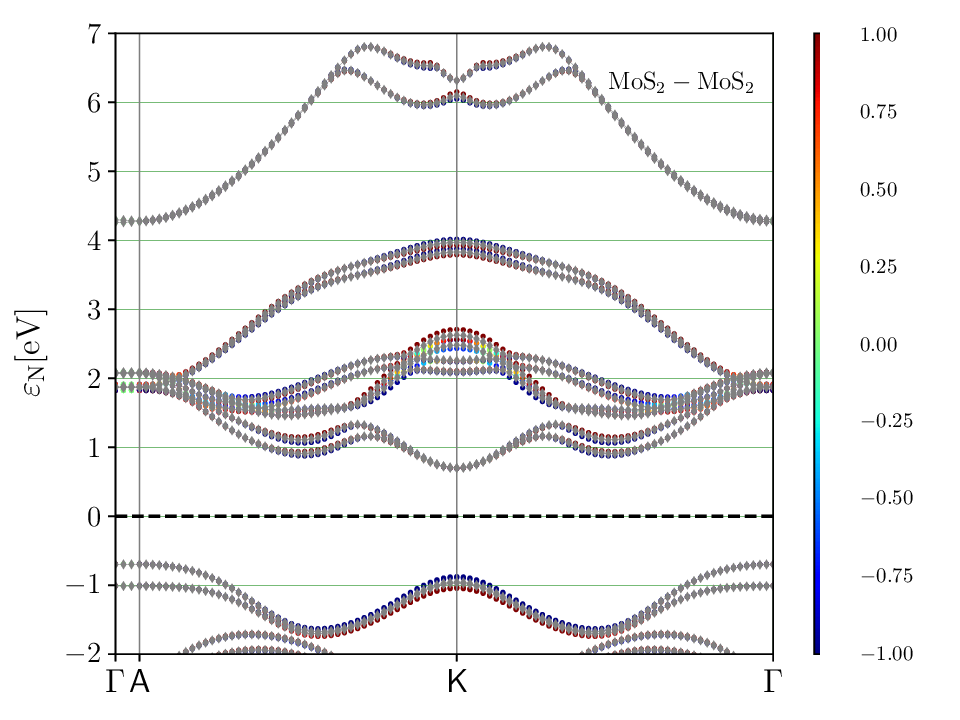}
\end{subfigure}%
\begin{subfigure}[t]{0.5\textwidth}
\centering
\includegraphics[width=\textwidth]{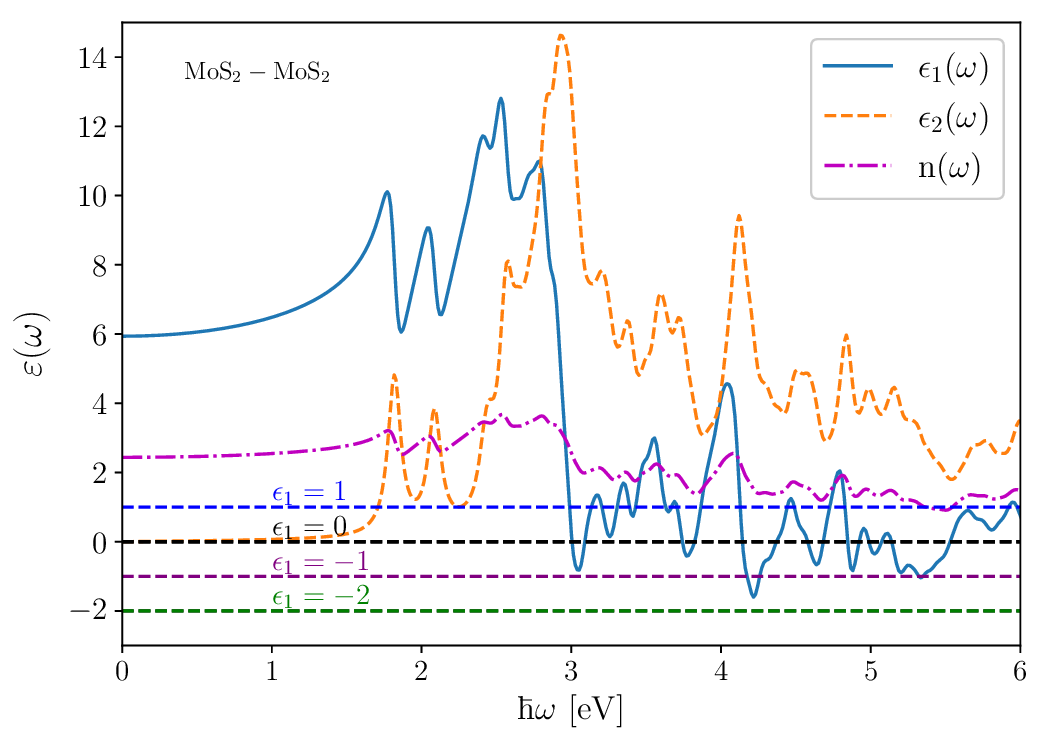}
\end{subfigure}
\begin{subfigure}[t]{0.5\textwidth}
\centering
\includegraphics[width=\textwidth]{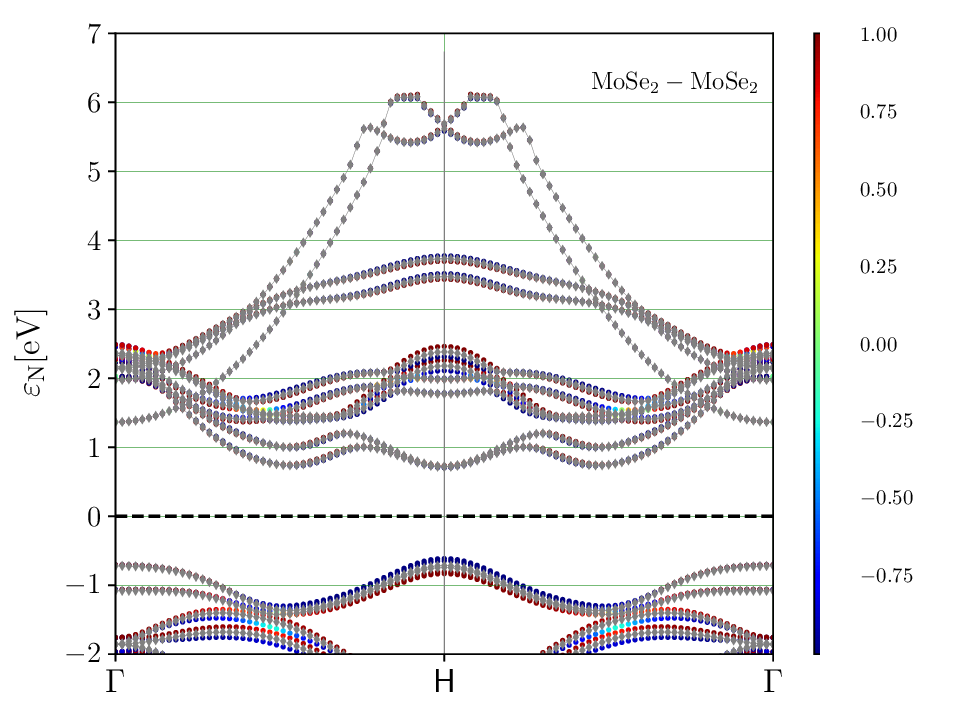}
\end{subfigure}%
\begin{subfigure}[t]{0.5\textwidth}
\centering
\includegraphics[width=\textwidth]{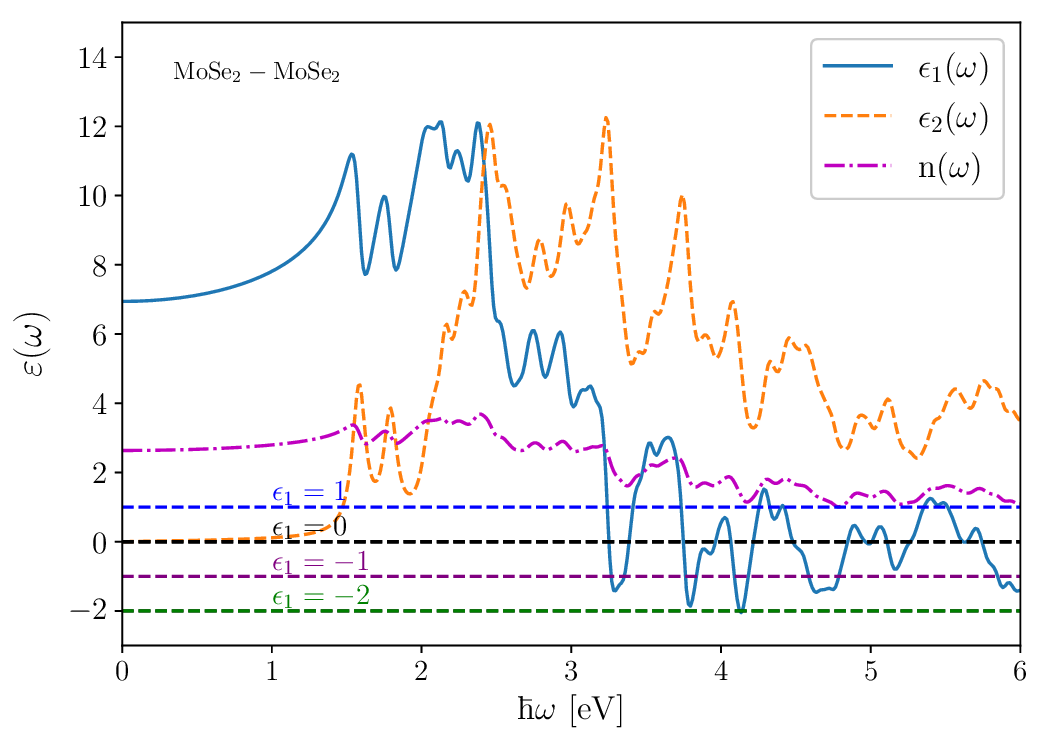}
\end{subfigure}
\caption{\label{fig22at1} Calculation of bilayer the $\varepsilon(k)$ connection with hybridization intensity is shown on the band structures, with the band, $\varepsilon_{N}$ at higher energy on the left and the dielectric function, $\mathrm{\epsilon(\omega)}$ and index of refraction, $\mathrm{n(\omega)}$ on the right.} 
\end{figure*}

\begin{figure*}[ht!]
\begin{subfigure}[t]{0.5\textwidth}
\centering
\begin{adjustbox}{max size={\textwidth}{\textheight}}
\includegraphics[scale=1]{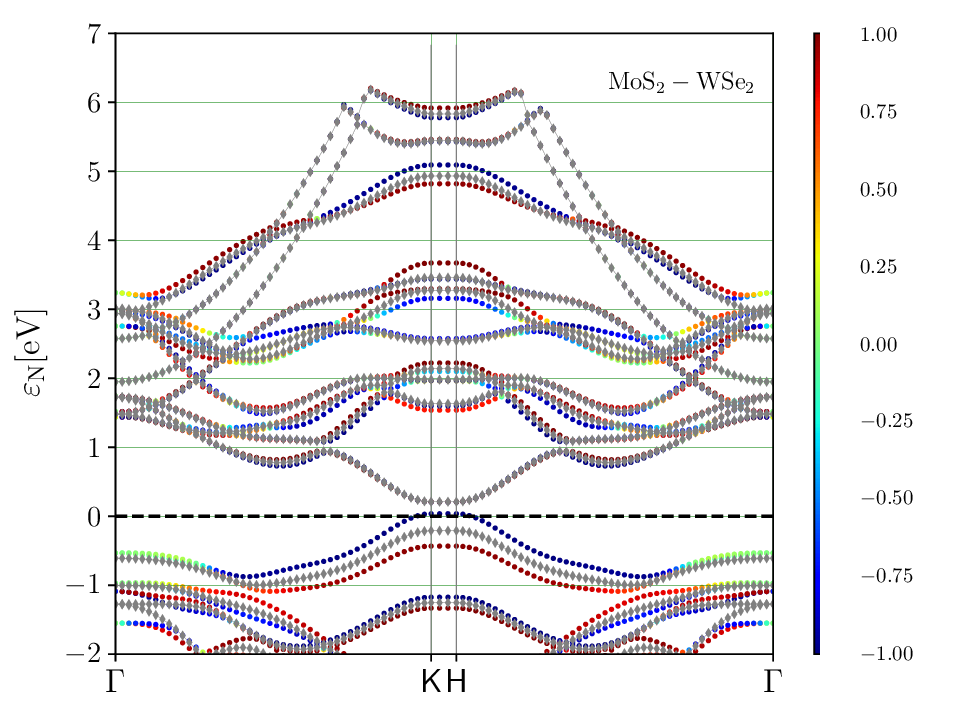}
\end{adjustbox}
\end{subfigure}%
\begin{subfigure}[t]{0.5\textwidth}
\centering
\begin{adjustbox}{max size={\textwidth}{\textheight}}
\includegraphics[scale=1]{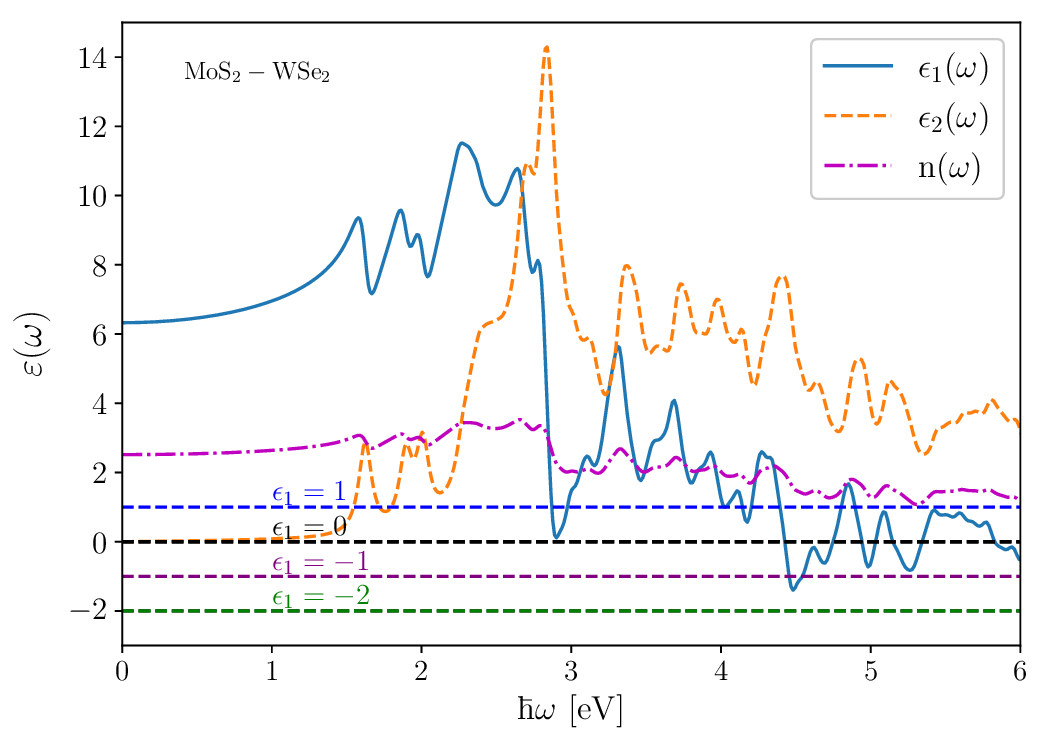}
\end{adjustbox}
\end{subfigure}
\begin{subfigure}[t]{0.5\textwidth}
\centering
\begin{adjustbox}{max size={\textwidth}{\textheight}}
\includegraphics[scale=1]{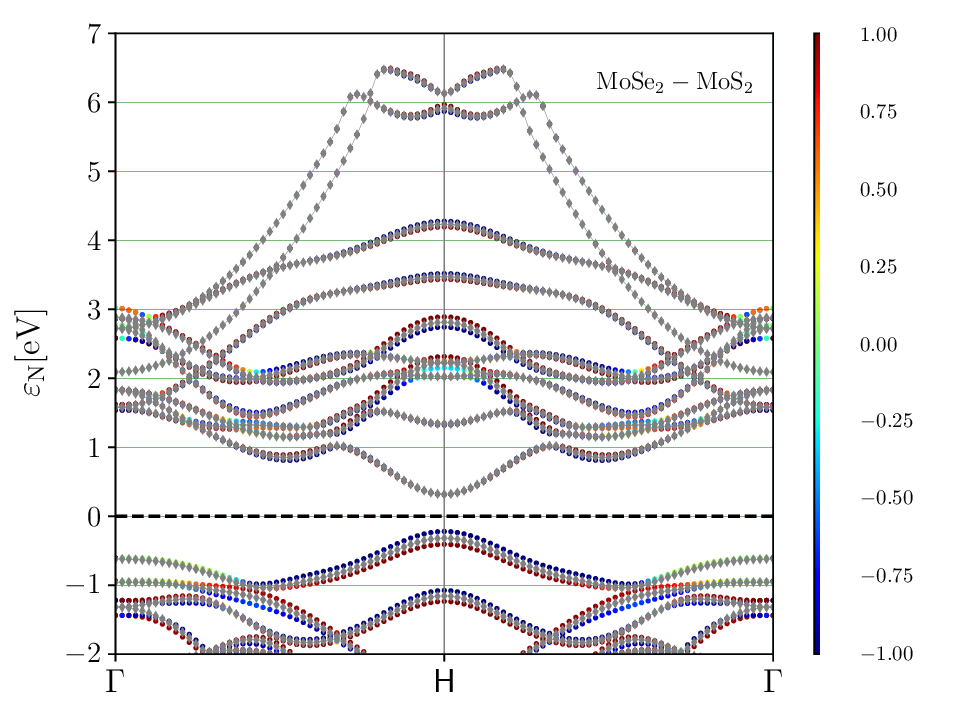}
\end{adjustbox}
\end{subfigure}%
\begin{subfigure}[t]{0.5\textwidth}
\centering
\begin{adjustbox}{max size={\textwidth}{\textheight}}
\includegraphics[scale=1]{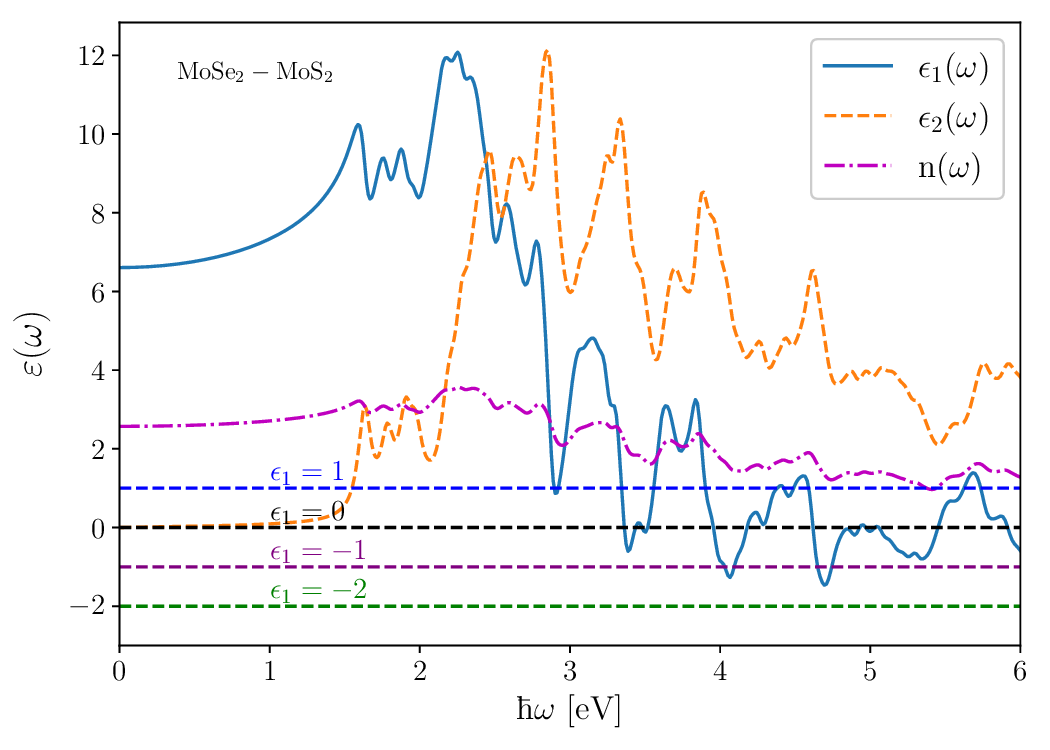}
\end{adjustbox}
\end{subfigure}
\begin{subfigure}[t]{0.5\textwidth}
\centering
\begin{adjustbox}{max size={\textwidth}{\textheight}}
\includegraphics[scale=1]{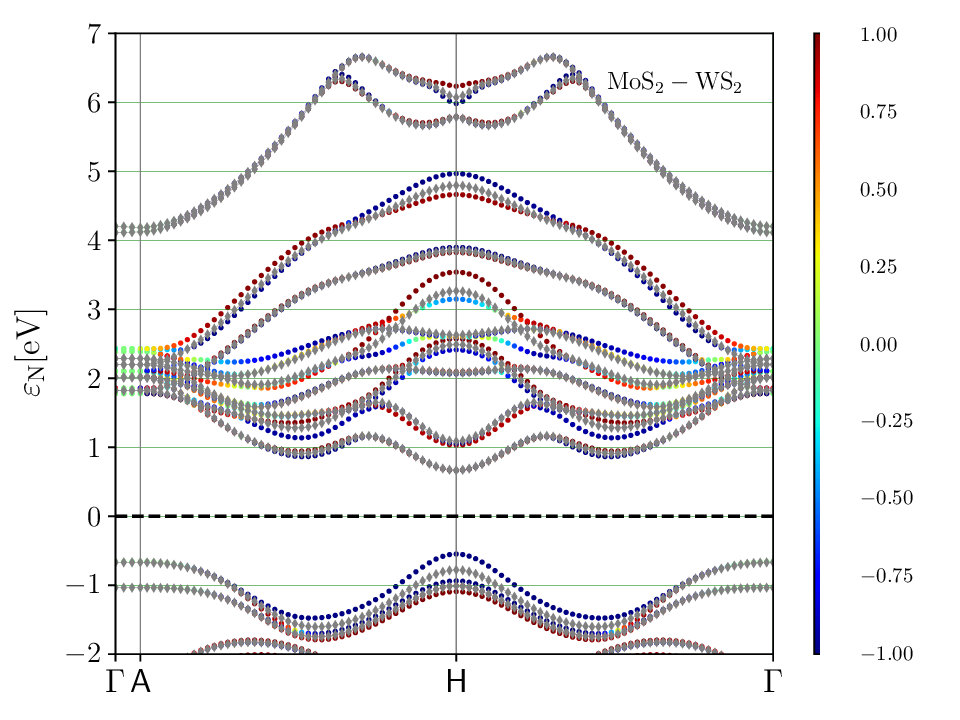}
\end{adjustbox}
\end{subfigure}%
\begin{subfigure}[t]{0.5\textwidth}
\centering
\begin{adjustbox}{max size={\textwidth}{\textheight}}
\includegraphics[scale=1]{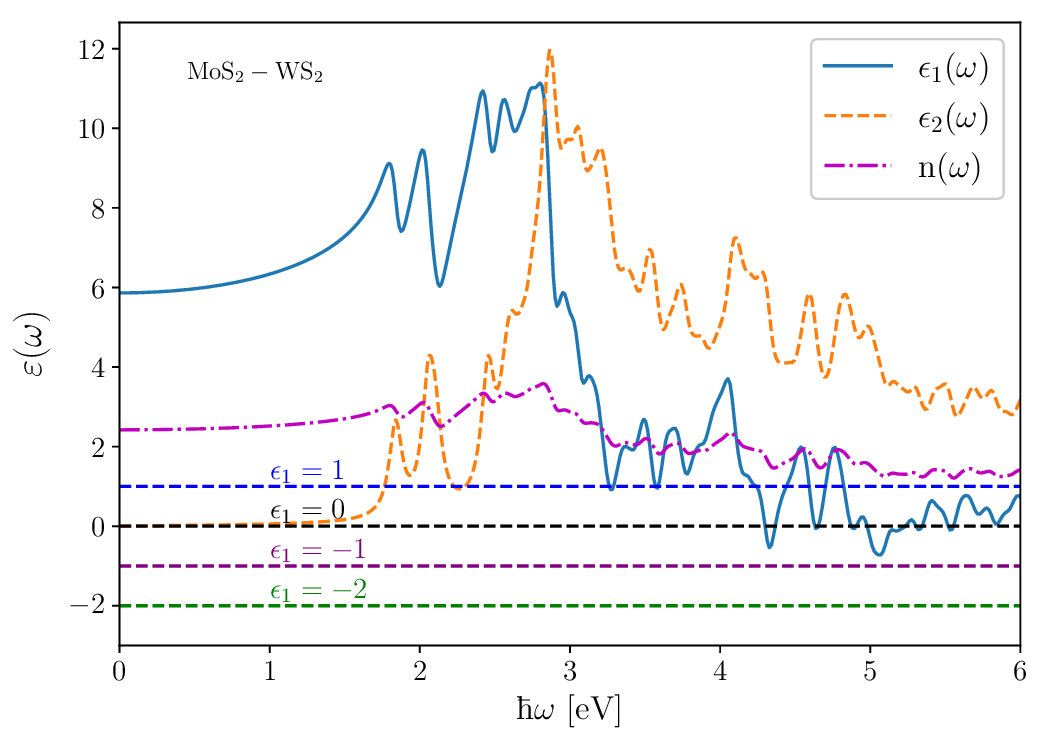}
\end{adjustbox}
\end{subfigure}
\caption{\label{fig23at1} Calculation of bilayer the $\varepsilon(k)$ connection with hybridization intensity is shown on the band structures, with the band, $\varepsilon_{N}$ at higher energy on the left and the dielectric function, $\mathrm{\epsilon(\omega)}$ and index of refraction, $\mathrm{n(\omega)}$  on the right.} 
\end{figure*}

Fig.~\ref{fig22at1} demonstrates that the negative dielectric constant has a significant effect on density of states. Plasmons or surface plasmon polariton-assisted excitation with $n\omega_{p}$ where n is any fraction resulted in a oscillation of cloud of hot electrons in the home structured configuration. Here hybridization has no significant use which one can notice from the band structure. $\mathrm{MoS_{2}-MoS_{2}}$ oscillates more evenly and may aid in electron excitation in the [$\frac{\omega_{p}}{\sqrt{3}}$, $\mathrm{\omega_{p}}$] vicinity than $\mathrm{MoSe_{2}-MoSe_{2}}$, which ranges from [$\frac{\omega_{p}}{\sqrt{3}}$, $\frac{\omega_{p}}{\sqrt{2}}$].

Fig~.\ref{fig23at1} shows that plasmons, surface plasmon-polaritons, and hybridization all enhance hot electron excitation while lowering thermal conductivity.  Hybridization, in particular, contributed greatly to the formation of several bands of the same energy, allowing hot electrons to relax and speed up the thermalization process. The $\mathrm{MoS_{2}-WS_{2}}$ surface promotes electron excitation with harmonic oscillation, leading to  more electrons. However, $\mathrm{MoSe_{2}-MoS_{2}}$ plasmons exit with disordered oscillation, resulting in fewer electrons, and the $\mathrm{MoS_{2}-WS_{2}}$ surface plays an important role in electron decay. 

\section{Conclusion\label{sec:conc}}
The tuning of 2D structures in various configurations is critical to structural, electronic, optical and thermoelectric properties. In this paper, we used  dispersion relation, $\mathrm{\varepsilon(k)}$  to calculate the effective mass and band curvature in connection to plausible layered structure and their impact on the density of hot electrons to the corresponding group velocity.
The effective mass for the same polarizibility can differ due to the interchanging layer layout, which is mostly caused by the band alignment effect. Hetro layered van der Waals with various sorts of surrounding layers increase intermolecular force between layers, resulting in a narrower electronic band gap and higher exciton binding energy. However, the identical layer structure has less interactions, resulting in a wider band gap and higher exciton binding energy. Furthermore, surface plasmon polariton-assisted electron cloud excitation is frequently observed in van der Waals heterostructures, and hybridization creates more unoccupied states for the thermal cooling of excited hot electrons. Layering in various configurations can so impact polarizability, effective mass, and density of states, resulting in distinct excitonic and thermoelectric properties of the material.

Therefore, it is essential to optimize the thermoelectric and excitonic properties of heterostructured solar materials in order to raise open source voltages and reduce heat conductivity, respectively. 

\newpage
\section*{Disclosure\ statement}
The author declare that there is no conflict of interest.

\section{Data\ Availability\ Statement}
The data that\ support the findings\ of\ 
this study\ are available\ upon reasonable\ 
request\ from the\ author.\
\section{Acknowledgements}
We are grateful to the Dilla University~for~financial~support. 
T.E.\ Ada.\\
\url{https://orcid.org/0000-0002-4417-0058}\\
K.N.\ Nigussa\\
\url{https://orcid.org/0000-0002-0065-4325}\\
Cecil.N.M\ Ouma\\
\url{https://orcid.org/0000-0001-7328-2254}\\
Eyasu Tadese.\ M\\
\url{https://orcid.org/0009-0008-1393-8440}\\
\section{References}
\bibliographystyle{elsarticle-num}
\bibliography{refs}
\end{document}